\documentclass[12pt]{article}
\begin{document}
\title{ Model of a relativistic oscillator in a generalized Schr\"odinger picture}
\date{}
\author{Rudolf A. Frick\thanks{Email: rf@thp.uni-koeln.de}\\ Institut f\"ur Theoretische Physik, Universit\"at K\"oln,\\ Z\"ulpicher Str. 77, 50937 K\"oln, Germany }  
\maketitle
\date{}
\begin{abstract}
In a generalized Schr\"odinger picture, we  consider the motion of a  relativistic particle in an external field (like in the case of a harmonic oscillator).  In this picture the analogs of the  Schr\"odinger operators  of the particle are independent of  both the time and the space coordinates. These operators  induce  operators which are related to Killing vectors of the Anti de Sitter (AdS) space. We also consider the nonrelativistic limit.  
\end{abstract}
\section{Introduction}
 In this paper we present a model of a  relativistic harmonic oscillator  which is based on the  expansions of the Lorentz group   and  a  generalized Schr\"odinger picture.  The functions that realize the unitary representation of the one-dimensional Lorentz group $(p$ = momentum, $m$ = mass, ${p^2_0-c^2p^2}=m^2c^4$) have the form 
\begin{equation}
\label{1.1}
{\xi}(p,{\alpha})={e}^{i\alpha\ln[(p_0-cp)/mc^2]},
\end{equation}
 and  are the eigenfunctions of the boost generator $N(p)=ip_0{\partial}_{cp}$, ($N{\Rightarrow}{\alpha}$). The following expansion (Shapiro transformation \cite{Shap}) of the wave function of a particle in the momentum representation ($\rho={\alpha}{\hbar}/mc$)
\begin{equation}
\label{1.2}
\psi({p})=\frac{1}{\sqrt{2\pi\hbar}}\int_{-\infty}^{^\infty}{\xi}(p,{\rho})\psi({\rho})d{\rho},
\end{equation}
leads to the functions $\psi({\rho})$ in the spacetime independent $\rho$-representation.   In \cite{Kad}, in the framework of a two-particle equation of the quasipotential type, the variable $\rho$ was interpreted as the relativistic generalization of a relative coordinate. Quasipotential models of a relativistic oscillator  were first considered in \cite{Donk,Atak,Atak2}.

The $\rho$-representation or  the $\rho{\bf n}$ representation ($\vert{\bf n}\vert=1)$  may  also be used in a so-called generalized Schr\"odinger (GS) picture in which the analogs of the Schr\"odinger operators of a particle are independent of both the time and the space coordinates t, ${\bf x}$ in different representations \cite{Fri1}. The one-dimensional motion  of a free particle in the $\rho$-representation is described by the equations
\begin{equation}
\label{1.3}
i\hbar\frac{\partial}{\partial{t}}\psi(\rho,t,x)=H(\rho)\psi(\rho,t,x),\quad{-i}\hbar\frac{\partial}{\partial{x}}\psi(\rho,t,x)=P(\rho)\psi(\rho,t,x),
\end{equation}
where the Hamilton operator $H(\rho)$ and the momentum operator $P(\rho)$ have the form
\begin{equation}
\label{1.4}
H({\rho})=mc^2\cosh(-\frac{i\hbar}{mc}{\partial}_{{\rho}}),{\quad}P({\rho})=mc\sinh(-\frac{i\hbar}{mc}{\partial}_{{\rho}}).
\end{equation}
 The operators $H(\rho)$, $P(\rho)$  and the generator of the Lorentz group $N(\rho)$  ($N(\rho)=\rho)$ satisfy the commutation relations of the Poincar\'e algebra,
\begin{equation}
\label{1.5}
\lbrack{N},P\rbrack=i\frac{\hbar}{mc^2}H,\quad\lbrack{P},{H}\rbrack=0,\quad\lbrack{H},{N}\rbrack=-i\frac{\hbar}{m}P.
\end{equation}
In the recent paper \cite{Fri2} it was shown that in the GS picture the propagators  with different $\rho$  in the relativistic region and nonrelativistic limit may be used to describe the motion of extended objects like strings. The main aim of the present paper is to describe  in this picture the  motion of a relativistic particle in an  external field like the harmonic oscillator potential.

 Our treatment is based on the following assumptions: We will introduce the operators $\hat{P_0}=H+H^{'}$ and $\hat{P_1}={P}+{P}^{'}$ instead of ${H}$ and ${P}$,  where $H^{'}$ and ${P}^{'}$ represent the external field in the ${p}$ or in the $\rho$-representations.  We make the assumption that only the operators  $\hat{P_0}$ and $\hat{P_1}$  contain the interaction parts. We assume that the equations of the motion of a particle in an external field may be written in the form of a direct generalization of the equations (\ref{1.3}),
\begin{equation}
\label{1.6}
K_0(t,x)\Psi=\hat{P_0}\Psi,\quad{K}_1(t,x)\Psi=\hat{P}_1\Psi.
\end{equation}
Here, in analogy to (\ref{1.3}) for the left-hand sides, we introduced operators $K_0(t,x)$ and $K_1(t,x)$ in terms of the spacetime coordinates $t,x$. In the present paper, taking the practically important example of the harmonic oscillator, we will show that the use of the equations (\ref{1.6}) makes it possible to give a description of the motion of a relativictic particle in an external field.  The external fields do violate the commutation relations of the Poincar\'e algebra, $\lbrack{\hat{P}_1},{\hat{P_0}}\rbrack\not=0$. We have the problem of determining  observables in  GS picture. In the equations (\ref{1.6}) the operators on the right-hand-side  are spacetime independent operators and thus correspond to constants of motion.  For  the harmonic relativistic oscillator we will show that  the equations (\ref{1.6}) induce  operators $K_0$, $K_1$ which are related to Killing vectors of the AdS space.  First we study the one-dimensional relativistic oscillator. We set up the rules which may be used to derive the explicit form of the operators $K_0$, ${K}_1$.  In Sect. 3 we will consider the nonrelativistic limit. In Sect. 4 the three-dimensional relativistic oscillator (spin = $0$) is considered.
\section{The one-dimensional relativistic oscillator} 
We consider the motion of a relativistic particle in  external field which is presented by the operators $(\omega$ = frequency)
\begin{equation}
\label{1.7}
H^{'}(\rho)=\frac{m{\omega}^2}{2}\rho\left(\rho-i\frac{\hbar}{mc}\right)e^{-i\frac{\hbar}{mc}{\partial}_{\rho}},{\quad}P_1^{'}(\rho)=\frac{m{\omega}^2}{2c}\rho\left(\rho-i\frac{\hbar}{mc}\right)e^{-i\frac{\hbar}{mc}{\partial}_{\rho}}. 
\end{equation}
For the operators  on the right-hand-side of the Eqs. (\ref{1.6}) we have
\begin{equation}
\label{1.8}
\hat{P_0}(\rho)={mc^2}\cosh\left(-\frac{i\hbar}{mc}{\partial}_{{\rho}}\right)+\frac{m{\omega}^2}{2}\rho\left(\rho-i\frac{\hbar}{mc}\right)e^{-i\frac{\hbar}{mc}{\partial}_{\rho}},
\end{equation}
\begin{equation}
\label{1.9}
\hat{P_1}(\rho)={mc}\sinh\left(-\frac{i\hbar}{mc}{\partial}_{{\rho}}\right)+\frac{m{\omega}^2}{2c}\rho\left(\rho-i\frac{\hbar}{mc}\right)e^{-i\frac{\hbar}{mc}{\partial}_{\rho}}.
\end{equation}
 The realizations of the operators $\hat{P_0}$ and $\hat{P_1}$  in  momentum representation are given in  appendix A.
In the nonrelativistic limit the operators $\hat{P_0}(\rho)-mc^2$ and $\hat{P_1}(\rho)$ assume the form
\begin{equation}
\label{1.10}
\hat{P_0}_{\rm{nr}}=-\frac{{\hbar}^2}{2m}\frac{{\partial}^2}{\partial{\rho}^2}+\frac{m{\omega}^2}{2}{\rho}^2,{\quad}\hat{P_1}_{\rm{nr}}=-i\hbar\frac{\partial}{\partial\rho}.
\end{equation} 
The operators $\hat{P_0}(\rho)$, $\hat{P_1}(\rho)$ and $\rho$ satisfy the commutation relations 
\begin{equation}
\label{1.11}
\lbrack\rho,\hat{P_1}\rbrack=i\frac{\hbar}{mc^2}\hat{P_0},\quad\lbrack\hat{P_1},\hat{P_0}\rbrack=-i\hbar{m}{\omega}^2\rho,\quad\lbrack\hat{P_0},\rho\rbrack=-i\frac{\hbar}{m}\hat{P_1}.
\end{equation}
The Casimir operator is a multiple of the identity operator ${I}$,
\begin{equation}
\label{1.12}
C(\rho)=\frac{1}{(\hbar\omega)^2}{\lbrace}\hat{P_0}^{2}-c^2\hat{P}_1^{2}{\rbrace}-\frac{m^2c^2}{{\hbar}^2}{\rho}^2=\left(\frac{mc^2}{\hbar\omega}\right)^2{I}.
\end{equation}

Let us examine the restrictions imposed on the operators $K_0(t,x)$ and $K_1(t,x)$ by the equations
\begin{equation}
\label{1.13}
K_0(t,x)\Psi(\rho;t,x)=\hat{P_0}(\rho){\Psi}(\rho;t,x),
\end{equation}
\begin{equation}
\label{1.14}
{K}_1(t,x){\Psi}(\rho;t,x)=\hat{P}_1(\rho){\Psi}(\rho;t,x),
\end{equation}
and the commutations rules (\ref{1.11}). We multiply the  equation (\ref{1.13}) from the left by the operator $\hat{P_1}({\rho})$, the equation (\ref{1.14}) from the left by $\hat{P_0}(\rho)$ and substract the two resulting equations one from the other. Bearing in mind that
\begin{equation}
\label{1.15}
 [\hat{P_1}({\rho}),K_0(t,x)]=0,\quad [\hat{P_0}({\rho}),K_1(t,x)]=0,
\end{equation}
 we find 
\begin{equation}
\label{1.16}
[K_0(t,x),K_1(t,x)]{\Psi}=-[\hat{P_0}({\rho}),\hat{P_1}({\rho})]{\Psi},
\end{equation}
or 
\begin{equation}
\label{1.17}
[K_1(t,x),K_0(t,x)]{\Psi}=i\hbar{m}{\omega}^2{\rho}{\Psi}.
\end{equation}
In the following we introduce the operator 
\begin{equation}
\label{1.18}
K(t,x)=\frac{1}{i\hbar{m}{\omega}^2}[K_1(t,x),K_0(t,x)]
\end{equation}
for which we have
\begin{equation}
\label{1.19}
K(t,x){\Psi}(\rho;t,x)={\rho}{\Psi}(\rho;t,x).
\end{equation}
Eqs. (\ref{1.16}) and 
\begin{equation}
\label{1.20}
[K_1,K]{\Psi}=-[\hat{P_1},{\rho}]{\Psi},\quad[K,K_0]{\Psi}=-[\rho,\hat{P_0}]{\Psi}
\end{equation}
state that the equations (\ref{1.13}),  (\ref{1.14}) and  (\ref{1.19}) induce operators $K_0$, $K_1$ and $K$  with the same commutation rules as  $\hat{P_0}$, $\hat{P_1}$ and $\rho$, except for the minus signs on the right-hand-sides 
\begin{equation}
\label{1.21}
\lbrack{K},{K_1}\rbrack=-\imath\frac{\hbar}{mc^2}K_0,\quad\lbrack{K_1},K_0\rbrack=i\hbar{m}{\omega}^2K,\quad\lbrack{K_0},K\rbrack=\imath\frac{\hbar}{m}K_1.
\end{equation}
If  at first we introduce commutation relations like (\ref{1.21}) for the operators $\hat{P_0}$, $\hat{P_1}$ and $\rho$, then  for $K_0$, $K_1$ and $K$  we must introduce commutation relations like (\ref{1.11}). This can be  made by replacing  $\hat{P_1}$ by -$\hat{P_1}$ and ${K_1}$ by -${K_1}$, respectively.
For the Casimir operators (\ref{1.12}) and  
\begin{equation}
\label{1.22}
C(t,x)=\frac{1}{(\hbar\omega)^2}{\lbrace}{K_0}^{2}-c^2{K}_1^{2}{\rbrace}-\frac{m^2c^2}{{\hbar}^2}{K}^2,
\end{equation}
we have the equation
\begin{equation}
\label{1.23}
C(t,x){\Psi}(\rho;t,x)=C(\rho){\Psi}(\rho;t,x).
\end{equation}

The operators $N_1=mcK/{\hbar}$, $N_2=cK_1/{\hbar\omega}$, $N_3=K_0/{\hbar\omega}$  satisfy the  commutation relations of the noncompact   Lie algebra $so(2,1),$
\begin{equation}
\label{1.24}
\lbrack{N}_1,N_2\rbrack=-iN_3,\quad\lbrack{N_2},N_3\rbrack=iN_1,\quad\lbrack{N_3},N_1\rbrack=iN_2.
\end{equation}
In the representations in which $N_1$ or $N_2$ is diagonal, their eigenvalue spectrum is continuous.  We are interested in a  representation in which $N_3$ is diagonal.  In this case, the eigenvalue spectrum of $N_3$ is discrete for an irreducible representation  and has the form of $N_3$ ${\longrightarrow}-a+n{\quad}(n=0,1,2,...)$, where the number $a$ is related to eigenvalues of the Casimir operator $C{\longrightarrow}{a}(a+1)$. For the unitary representations ($D^{+}$-series), $-a=1/2,1,3/2,2,...$ \cite{Bar}. It follows from (\ref{1.12}) and (\ref{1.23}) that ${a}=-1/2-\sqrt{1/4+({mc^2}/{\hbar\omega})^2}$. The cases $-a=1/2$ and $-a=1$ must be rejected. For the cases $-a=n_0=3/2,2,5/2,$.., we have  ${K_0}{\longrightarrow}\hbar\omega(n+1/2+\mu)$, where $\mu=\sqrt{1/4+({mc^2}/{\hbar\omega})^2}$.

 Similarly, putting  $N_1=mc\rho/{\hbar}$, $N_2=-c\hat{P}_1/{\hbar\omega}$, $N_3=\hat{P}_0/{\hbar\omega}$, we find that in the  representation in which the  operator $\hat{P_0}(\rho)$ is diagonal, its  spectrum  is $E_{n\mu}=\hbar\omega(n+1/2+\mu)$. For $mc^2/{\hbar\omega}\gg{1/2}$ we have $E_{n\mu}\approx{\hbar\omega}(n+1/2)+mc^2$. 

For convenience, in the text below, we omit to mention explicitly the number $n_{0}$  (except of Eqs. (\ref{1.43})-(\ref{1.44})). 

The explicit forms of the operators $K_0$, $K_1$ and $K$ depend on the realisation in terms of the spacetime coordinates. In order to interpret the operator $\hat{P_0}$ as  Hamilton operator, we chose the following realisation
\begin{equation}
\label{1.25}
K_0=i\hbar\frac{\partial}{\partial{t}},
\end{equation}
\begin{equation}
\label{1.26}
K_1=\sqrt{1+(\omega{x}/c)^2}\cos\omega{t}({i\hbar\partial}_{x})-\frac{(\omega{x/c^2})\sin\omega{t}}{\sqrt{1+(\omega{x}/c)^2}}i{\hbar\partial}_{t},
\end{equation}
\begin{equation}
\label{1.27}
K=\frac{x\cos\omega{t}}{mc^2\sqrt{1+(\omega{x/c})^2}}i{\hbar\partial}_{t}+\frac{1}{m\omega}\sqrt{1+(\omega{x/c)^2}}\sin\omega{t}({i\hbar\partial}_{x}).
\end{equation}
We have the equation
\begin{equation}
\label{1.28}
i\hbar\frac{\partial}{\partial{t}}\Psi(\rho;t,x)=\hat{P_0}(\rho){\Psi}(\rho;t,x),
\end{equation}
which defines the operator $\hat{P_0}(\rho)$ as Hamilton operator.   The operators (\ref{1.25}) to (\ref{1.27}) are related to  Killing vectors of the AdS space with metric
\begin{equation}
\label{1.29}
ds^2=\left(1+\frac{{\omega}^2x^2}{c^2}\right)c^2dt^2-\frac{1}{1+\frac{{\omega}^2x^2}{c^2}}dx^2.
\end{equation}
 A general solution of $\Psi(\rho;t,x)$  can be written as a sum of separated solutions 
\begin{equation}
\label{1.30}
\Psi(\rho;t,x)=\sum_{n=0}^{\infty}v_n(\rho)f_n(t,x),
\end{equation}
\begin{equation}
\label{1.31}
i\hbar\frac{\partial}{\partial{t}}v_n(\rho)f_n(t,x)=\hat{P_0}(\rho)v_n(\rho)f_n(t,x),
\end{equation}
where $v_n(\rho)$ are the eigenfunctions of the operator $\hat{P}_0(\rho)$ and $f_n(t,x)$ are the eigenfunctions of the operators $K_0$ and $C(t,x)$. In the following we express the functions  $v_n(\rho)f_n(t,x)$ in terms of the operators 
\begin{equation}
\label{1.32}
A^{+}=\frac{1}{\sqrt{2}}\left\lbrace{\sqrt\frac{m\omega}{\hbar}}\rho-\frac{i}{\sqrt{m\hbar\omega}}\hat{P_1}\right\rbrace,{\quad}K^{+}=\frac{1}{\sqrt{2}}\left\lbrace{\sqrt\frac{m\omega}{\hbar}}K+\frac{i}{\sqrt{m\hbar\omega}}K_1\right\rbrace
\end{equation}
and the  functions ($\tau={\omega}t$, $y=\omega{x}/c$, $l=c/\omega$)
\begin{equation}
\label{1.33}
{v}_0(\rho)=2^{\mu}\sqrt{\frac{1}{\pi\Gamma(2\mu+1)}}\left(\frac{\hbar\omega}{mc^2}\right)^{-i\frac{mc}{\hbar}\rho}{\Gamma}\left(\mu+1/2-i\frac{mc}{\hbar}\rho\right), 
\end{equation}
\begin{equation}
\label{1.34}
{f}_0(\tau,y)=\sqrt\frac{\Gamma(\mu+\frac{1}{2})}{{\sqrt\pi}\Gamma(\mu+1)}e^{-i({\mu+\frac{1}{2}})\tau}{(1+y^2)^{-\frac{\mu+\frac{1}{2}}{2}}}.
\end{equation}
The functions ${v}_0$ and  ${f}_0$ satisfy the equations
\begin{equation}
\label{1.35}
A^{-}{v}_0(\rho)=0,{\quad}K^{-}{f}_0(\tau,y)=0,
\end{equation}
where
\begin{equation}
\label{1.36}
A^{-}=\frac{1}{\sqrt{2}}\left\lbrace{\sqrt\frac{m\omega}{\hbar}}\rho+\frac{i}{\sqrt{m\hbar\omega}}\hat{P_1}\right\rbrace,{\quad}K^{-}=\frac{1}{\sqrt{2}}\left\lbrace{\sqrt\frac{m\omega}{\hbar}}K-\frac{i}{\sqrt{m\hbar\omega}}K_1\right\rbrace.
\end{equation}

Using the operators $A^{+}$ and $K^{+}$, we can construct a system of normalized functions $(b=\sqrt{\frac{\hbar\omega}{2mc^2}})$
\begin{equation}
\label{1.37}
{v}_n(\rho)={\beta}_n(A^{+})^{n}{v}_0(\rho),{\quad}{f}_n(\tau,y)={\beta}_n(K^{+})^{n}{f}_0(\tau,y),
\end{equation}
where 
\begin{equation}
\label{1.38}
{\beta}_n={b}^{-n}\sqrt{\frac{\Gamma(2\mu+1)}{n!\Gamma(n+2\mu+1)}}.
\end{equation}
The explicit forms of the functions ${v}_n(\rho)$ and ${f}_n(\tau,y)$  ($n$ = 1,2,3... .) are given in  appendix B.
For the functions   ${\phi}_n(\rho,\tau,y)={v}_n(\rho){f}_n(\tau,y)$, we have the relations
\begin{equation}
\label{1.39}
A^{-}K^{-}{\phi}_n(\rho,\tau,y)=g^{2}_n{\phi}_{n-1}(\rho,\tau,y),
\end{equation}
\begin{equation}
\label{1.40}
{\quad}A^{+}K^{+}{\phi}_n(\rho,\tau,y)=g^{2}_{n+1}{\phi}_{n+1}(\rho,\tau,y),
\end{equation}
where $g_n={b}\sqrt{n(n+2\mu)}$.

Additionally,
\begin{equation}
\label{1.41}
A^{+}\Psi(\rho;t,x)=K^{-}\Psi(\rho;t,x),{\quad}A^{-}\Psi(\rho;t,x)=K^{+}\Psi(\rho;t,x).
\end{equation}
For the propagator we have the expression
\begin{equation}
\label{1.42}
{\cal K}(2,1)=\sum_{n=0}^{\infty}{\phi}_n({\rho_2},{t_2},x_2){\phi}^{*}_n(\rho_1,t_1,x_1).
\end{equation}
 
 The expression for the energy levels $E_{n\mu}=\hbar\omega(n+1/2+\mu)$ can be written in the form
\begin{equation}
\label{1.43}
E_{n{n_0}}= \hbar\omega(n+n_0)=\frac{mc^2}{\sqrt{n_0(n_0-1)}}(n+n_0).
\end{equation}
This shows that the oscillator frequency is discrete and for higher $n_0$ decreases accoding to
\begin{equation}
\label{1.44}
\omega\sim\frac{mc^2}{\hbar\sqrt{n_0(n_0-1)}}.
\end{equation}
The set of the numbers $n_0$ is infinite. For $n_0\to\infty$, one has $\omega\to{0}$. For the energy levels $E_{0{n_0}\rightarrow\infty}-mc^2\to{0}$. There is no zero-point energy. We cannot introduce the notion of a ground state like the ground state of the nonrelativistic quantum oscillator in quantum mechanics.

For the AdS radius $\kappa=c/{\omega}$, we have $\kappa=\sqrt{n_0(n_0-1)}$ $\hbar/{mc}$.
\section{The nonrelativistic problem} 
In the nonrelativistic limit, the operators (\ref{1.10}) and $\rho$ satisfy the commutation relations 
\begin{equation}
\label{1.45}
\lbrack{\rho},\hat{P_1}_{\rm{nr}}\rbrack=\imath\hbar,\quad\lbrack{\hat{P_1}}_{\rm{nr}},\hat{P_0}_{\rm{nr}}\rbrack=-i\hbar{m}{\omega}^2{\rho},\quad\lbrack{\hat{P_0}}_{\rm{nr}},{\rho}\rbrack=-\imath\frac{\hbar}{m}\hat{P_1}_{\rm{nr}}.
\end{equation}
In the operators $K(t,x)$ and ${K_1}(t,x)$ we replace  the operator  $i{\hbar\partial}_{t}$ by $mc^2$ and assume that $({\omega}x/c)^2<1$. In this case the operators 
\begin{equation}
\label{1.46}
{K_0}_{\rm{nr}}(t,x)=K_0(t,x)=i\hbar\frac{\partial}{\partial{t}},{\quad}K_{\rm{nr}}(t,x)=x\cos{\omega}t+\frac{1}{m\omega}\sin\omega{t}({i\hbar\partial}_{x}),
\end{equation}
\begin{equation}
\label{1.47}
{K_1}_{\rm{nr}}(t,x)=-xm\omega\sin{\omega}t+\cos\omega{t}({i\hbar\partial}_{x}),
\end{equation}
satisfy  the commutation relations  
\begin{equation}
\label{1.48}
[K_{\rm{nr}},{K_1}_{\rm{nr}}]=-i\hbar,{\quad}[{K_1}_{\rm{nr}},K_0]=i\hbar{m}{\omega}^{2}K_{\rm{nr}},{\quad}[{K_0},K_{\rm{nr}}]=i\frac{\hbar}{m}{K_1}_{\rm{nr}}.
\end{equation}
For the Casimir relation we have 
\begin{equation}
\label{1.49}
C_{\rm{nr}}(t,x)=K_0-\frac{1}{2m}{K^2_1}_{\rm{nr}}-\frac{m{\omega}^2}{2}K^2_{\rm{nr}}=i\hbar\frac{\partial}{\partial{t}}-\widetilde{H}(x),
\end{equation}
where the operator $\widetilde{H}(x)$ have the form of the harmonic-oscillator Hamilton operator in quantum mechanics,
\begin{equation}
\label{1.50}
\widetilde{H}(x)=-\frac{{\hbar}^2}{2m}\frac{{\partial}^2}{\partial {x}^2}+\frac{m{\omega}^2x^2}{2}.
\end{equation}
The  Hamilton operator  $\hat{P_0}_{\rm{nr}}(\rho)$ in the GS picture has the same form as the operator $\widetilde{H}(x)$.

In the nonrelativistic limit, the operators $A^{+}({\rho})$, $K^{+}({t,x})$,  $A^{-}({\rho})$, and  $K^{-}(t,x)$ go over into  the operators ($\widetilde\rho=\sqrt\frac{m\omega}{\hbar}\rho$, $\widetilde{x}=\sqrt\frac{m\omega}{\hbar}{x}$)
\begin{equation}
\label{1.51}
a^{+}(\widetilde\rho)=\frac{1}{\sqrt{2}}\lbrace\widetilde\rho-\partial_{\widetilde\rho}\rbrace,{\quad}k^{+}(t,\widetilde{x})=e^{-i{\omega}t}\frac{1}{\sqrt{2}}\lbrace\widetilde{x}-\partial_{\widetilde{x}}\rbrace ,
\end{equation}
and
\begin{equation}
\label{1.52}
a^{-}({\widetilde\rho})=\frac{1}{\sqrt{2}}\lbrace\widetilde\rho+\partial_{\widetilde\rho}\rbrace,{\quad}k^{-}(t,\widetilde{x})=e^{i{\omega}t}\frac{1}{\sqrt{2}}\lbrace\widetilde{x}+\partial_{\widetilde{x}}\rbrace ,
\end{equation}
respectively.  

 The  solution of the equation 
\begin{equation}
\label{1.53}
i\hbar\frac{\partial}{\partial{t}}\Psi(\rho;t,x)_{\rm{nr}}=\hat{P_0}_{\rm{nr}}(\rho){\Psi}(\rho;t,x)_{\rm{nr}}
\end{equation}
can be written as a sum of separate solutions 
\begin{equation}
\label{1.54}
i\hbar\frac{\partial}{\partial{t}}w_n(\widetilde\rho)u_n(t,\widetilde{x})=\hat{P_0}_{\rm{nr}}(\widetilde\rho)w_n(\widetilde\rho)u_n(t,\widetilde{x}),
\end{equation}
where
\begin{equation}
\label{1.55}
{w}_n(\widetilde\rho)u_n(t,\widetilde{x})={\frac{1}{n!}}(a^{+})^{n}{w}_0(\widetilde\rho) (k^{+})^{n}u_0(t,\widetilde{x}).
\end{equation}
Here, the functions $(N_0=(m\omega/{\pi\hbar})^{1/4})$
\begin{equation}
\label{1.56}
w_0(\widetilde\rho)=N_0e^{-{\widetilde\rho}^2/2},{\quad}u_0(t,\widetilde{x})=N_0e^{-i{E_0}t/\hbar}e^{-{\widetilde{x}}^2/2},
\end{equation}
satisfy the equations
\begin{equation}
\label{1.57}
a^{-}{w}_0(\widetilde\rho)=0,{\quad}k^{-}{u}_0(t,\widetilde{x})=0.
\end{equation}
As a result, we have
\begin{equation}
\label{1.58}
{\Psi}(\rho;t,x)_{nr}=\sum_{n=0}^{\infty}N_ne^{-{\widetilde\rho}^2/2}H_n(\widetilde\rho){N_n}e^{-i{E_n}t/\hbar}e^{-{\widetilde{x}}^2/2}H_n(\widetilde{x}),
\end{equation}
where
\begin{equation}
\label{1.59}
u_n(t,\widetilde{x})=N_ne^{-i{E_n}t/\hbar}e^{-{\widetilde{x}}^2/2}H_n(\widetilde{x})
\end{equation}
are the well-known harmonic-oscillator wave functions in quantum mechanics.
The spectrum of the operators $\hat{P_0}_{\rm{nr}}(\rho)$ is $E_n=\hbar\omega(n+1/2)$.
For the propagator  we have the expression 
\begin{equation}
\label{1.60}
{\cal K}(2,1)_{nr}=\sum_{n=0}^{\infty}w_n(\rho_2)u_n(t_2,{x_2})w^{*}_n(\rho_1)u^{*}_n(t_1,{x_1}).
\end{equation}

\section{The three-dimensional relativistic oscillator $(\rm{spin=0})$}

 The motion  of a free particle whit spin $0$ is described by the equations ($0\leq\rho<\infty$, ${\bf n}=(\sin{\theta_1}\cos{\varphi_1},\sin{\theta_1}\sin{\varphi_1},\cos{\theta_1}))$
\begin{equation}
\label{1.61}
i\hbar\frac{\partial}{{\partial}t}\psi(\rho,{\bf n},t,{\bf x})=H\psi(\rho,{\bf n},t,{\bf x});\quad{-i}\hbar\frac{\partial}{{\partial}{\bf x}}\psi(\rho,{\bf n},t,{\bf x})={\bf P}\psi(\rho,{\bf n},t,{\bf x}),
\end{equation}
where $H(\rho,{\bf n})$ and  ${\bf P}(\rho,{\bf n})$ are the Hamilton and the momentum operators of the particle. They  have the form
\begin{equation}
\label{1.62}
H(\rho,{\bf n})=mc^2\cosh\left(\frac{i\hbar}{mc}{\partial}_{\rho}\right)+\frac{i\hbar{c}}{{\rho}}{\sinh}\left(\frac{i\hbar}{mc}{\partial}_{\rho}\right) +\frac{{\bf L}^2({\bf n})}{2m{\rho}^2}e^{\frac{i\hbar}{mc}{\partial}_{\rho}},
\end{equation} 

\begin{equation}
\label{1.63}
{\bf P}={\bf n}H/c-\frac{mc}{\rho}e^{{\frac{i\hbar}{mc}{\partial}_{\rho}}}{\bf N}(\rho,{\bf n}).
\end{equation}
Here ${\bf L}({\bf n})$ and ${\bf N}(\rho,{\bf n})={\rho}{\bf n}+({\bf n}\times{\bf L}-{\bf L}\times{\bf n})/2mc,$ are the operators of the Lorentz algebra in the $\rho{\bf n}$-representaion.

For the particle in an  external field like the tree-dimensional harmonic oscillator potential we use  the following operators
\begin{equation}
\label{1.64}
\hat{P_0}(\rho,{\bf n})=H(\rho,{\bf n})+H^{'}_0(\rho,{\bf n}),
\end{equation}
\begin{equation}
\label{1.65}
\hat{P}_i(\rho,{\bf n})=P_i(\rho,{\bf n})+P^{'}_i(\rho,{\bf n}),
\end{equation}
where
\begin{equation}
\label{1.66}
H^{'}_0=\frac{m{\omega}^2}{2}\left(\rho-i\frac{\hbar}{mc}\right)^2e^{-i\frac{\hbar}{mc}{\partial}_{\rho}},{\quad}P^{'}_i= n_i\frac{m{\omega}^2}{2c}\left(\rho-i\frac{\hbar}{mc}\right)^2e^{-i\frac{\hbar}{mc}{\partial}_{\rho}}.
\end{equation}
In the nonrelativistic limit 
\begin{equation}
\label{1.67}
 \hat{P_0}(\rho,{\bf n})-mc^2\longrightarrow-\frac{{\hbar}^2}{{2m}{\rho}^2}\frac{\partial}{{\partial}{\rho}}{\rho}^{2}\frac{\partial}{{\partial}{\rho}}+\frac{\bf L^{2}({\bf n})}{2m{\rho}^2}+\frac{m{\omega}^2}{2}{\rho}^2.
\end{equation}

The operators $\hat{P_0}(\rho,{\bf n})$, $\hat{P}_i(\rho,{\bf n})$, and ${\bf L}({\bf n})$, ${\bf N}(\rho,{\bf n})$ satisfy the  commutations rules 
\begin{equation}
\label{1.68}
\lbrack{N_i},\hat{P}_j\rbrack=\frac{\imath\hbar}{mc^2}\delta_{ij}\hat{P_0},\quad\lbrack\hat{P}_i,\hat{P_0}\rbrack=-\imath\hbar{m}{\omega}^2{N_i}\quad\lbrack\hat{P_0},{N_i}\rbrack=-\frac{\imath\hbar}{m}\hat{P_i},
\end{equation}
\begin{equation}
\label{1.69}
\lbrack\hat{P}_i,\hat{P}_j\rbrack=-\imath\hbar\frac{{\omega}^2}{c^2}\epsilon_{ijk}{L_k},\quad\lbrack{L_i},\hat{P_0}\rbrack=0,\quad\lbrack\hat{P}_i,L_j\rbrack=\imath\hbar\epsilon_{ijk}\hat{P}_k,
\end{equation}
\begin{equation}
\label{1.70}
\lbrack{L_i},{L_j}\rbrack=\imath\hbar\epsilon_{ijk}{L_k},\quad
\lbrack{N_i},{N_j}\rbrack=-\frac{\imath\hbar}{m^2c^2}\epsilon_{ijk}{L_k},\quad\lbrack{N_i},{L_j}\rbrack=\imath\hbar\epsilon_{ijk}{N_k}
\end{equation}
The set of the operators $\lbrace\hat{P_0}/{\hbar\omega},c\hat{P}_i/{\hbar\omega},\frac{mc}{\hbar}{N_i},\frac{1}{\hbar}{L_i}\rbrace$  form a  basis for the $\rho{\bf n}$-representation of the $SO(3,2)$ group generators.
The Casimir operator 
\begin{equation}
\label{1.71}
C(\rho,{\bf n})=\frac{1}{(\hbar\omega)^2}\left\lbrace\hat{P_0}^{2}-c^2\sum_{i=1}^{3}\hat{P}_i^{2}\right\rbrace-\frac{m^2c^2}{{\hbar}^2}{\bf N}^2+\frac{1}{{\hbar}^2}{\bf L}^2
\end{equation} 
is a multiple of the identity operator 
\begin{equation}
\label{1.72}
C(\rho,{\bf n})=\left(\frac{m^2c^4}{{\hbar}^2{\omega}^2}-2\right){I}.
\end{equation}
To proceed further we now introduce  the following ten operators in terms of the spacetime coordinates  ($x_1=r\sin\theta_2\cos{\varphi_2}$,  $x_2=r\sin\theta_2\sin\varphi_2$,  $x_3=r\cos\theta_2$), 
\begin{equation}
\label{1.73}
K_{04}=i\hbar\frac{\partial}{\partial{t}},
\end{equation}
\begin{equation}
\label{1.74}
K_{i4}=\sqrt{1+(\omega{r}/c)^2}\cos\omega{t}({i\hbar\partial}_{x_i})-\frac{(\omega{x_i/c^2})\sin\omega{t}}{\sqrt{1+(\omega{r}/c)^2}}i{\hbar\partial}_{t}.
\end{equation}
 \begin{equation}
\label{1.75}
K_{i0}=\frac{1}{m\omega}\sqrt{1+(\omega{r/c)^2}}\sin\omega{t}({i\hbar\partial}_{x_i})+\frac{x_i\cos\omega{t}}{mc^2\sqrt{1+(\omega{r/c})^2}}i{\hbar\partial}_{t},
\end{equation}
\begin{equation}
\label{1.76}
K_{ij}=i\hbar\left(x_i\frac{\partial}{\partial{x_j}}-x_j\frac{\partial}{\partial{x_i}}\right).
\end{equation}
 A direct calculation shows that the set of the operators $\lbrace{K_{04},K_{i4},K_{i0},K_{ij}}\rbrace$ determines the same Lie algebra  as the operators $\lbrace\hat{P_0},\hat{P}_i,N_i,L_i\rbrace$ except for the minus signs on the right-hand sides
\begin{equation}
\label{1.77}
\lbrack{K_{i0},K_{j4}}\rbrack=-\frac{\imath\hbar}{mc^2}\delta_{ij}K_{04},\quad\lbrack{K_{i4},K_{04}}\rbrack=\imath\hbar{m}{\omega}^2{K_{i0}},\quad\lbrack{K_{04},K_{i0}}\rbrack=\frac{\imath\hbar}{m}{K_{i4}},
\end{equation}
\begin{equation}
\label{1.78}
\lbrack{K}_{i4},{K}_{j4}\rbrack=\imath\hbar\frac{{\omega}^2}{c^2}K_{ij},\quad\lbrack{K_{ij},K_{04}}\rbrack=0,\quad\lbrack{K}_{i4},K_{ik}\rbrack=\imath\hbar{K}_{k4},
\end{equation}
\begin{equation}
\label{1.79}
\lbrack{K_{i0},K_{j0}}\rbrack=\frac{\imath\hbar}{m^2c^2}K_{ij},\quad\lbrack{K_{i0},K_{ik}}\rbrack=\imath\hbar{K_{k0}}.
\end{equation}
The operators $\lbrace{K_{04},K_{i4},K_{i0},K_{ij}}\rbrace$ are related to  Killing vectors of the  AdS space with metric
\begin{equation}
\label{1.80}
ds^2=\left(1+\frac{{\omega}^2r^2}{c^2}\right)c^2dt^2-\frac{1}{1+\frac{{\omega}^2r^2}{c^2}}dr^2-r^2(d{\theta_2}^2+\sin^2\theta_2{d{\varphi_2}^2}).
\end{equation}
Following a procedure  similar to that of Sect. 2, we can introduce the equation
\begin{equation}
\label{1.81}
i\hbar\frac{\partial}{\partial{t}}\Phi(\rho,{\bf n};t,{\bf x})=\hat{P_0}(\rho,{\bf n})\Phi(\rho,{\bf n};t,{\bf x}),
\end{equation}
which define the operator $\hat{P_0}(\rho,{\bf n})$ as Hamilton operator.
Additionally,
\begin{equation}
\label{1.82}
K_{i4}\Phi=\hat{P_i}\Phi,{\quad}K_{i0}\Phi= N_i\Phi,{\quad}K_{ij}\Phi= {\varepsilon}_{ijk}L_k\Phi.
\end{equation}
The Casimir operator like (\ref{1.71}) may be written in the form $(\tau={\omega}t$, ${\tan}\sigma={\omega}r/c$)
\begin{equation}
\label{1.83}
C({\tau},\sigma,\theta_2,\varphi_2)=-\left\lbrace{\cos^2\sigma}\frac{{\partial}^2}{\partial{{\tau}^2}}-{\cot^2\sigma\lbrack{\cos^2\sigma}\frac{\partial}{\partial{\sigma}}({\tan}^2\sigma\frac{\partial}{{\partial}\sigma})+{\Lambda}(\theta_2,\varphi_2)}\rbrack\right\rbrace.
\end{equation}

 The solutions  with  discrete  spectrum  are \cite{Avis} $(n=0,1,2...,$ $l=0,1,2...,$ $l\geq\vert{m}\vert$, $M=3,4,5...$, $\lambda=2n+l+M)$ 
\begin{equation}
\label{1.84}
{\psi}^{M}_{nlm}=N^{M}_{nl}e^{-i\lambda\tau}(\cos\sigma)^{M}(\sin\sigma)^{l}P^{(l+1/2,M-3/2)}_n({\cos}2\sigma)Y^{\ast}_{lm}(\theta_2,\varphi_2),
\end{equation}
where  $P^{(\alpha,\beta)}_n(x)$ are the Jacobi polynomials and $N^{M}_{\rm{nl}}$ are normalization constants  
\begin{equation}
\label{1.85}
N^{M}_{nl}=\left[\frac{n!\Gamma(M+l+n)}{\Gamma(M+n-1/2)\Gamma(l+n+3/2)}\right]^{1/2}.
\end{equation}
The number $M$ is related to the eigenvalues of the Casimir operator
\begin{equation}
\label{1.86}
C(\tau,\sigma,\theta_2,\varphi_2){\Longrightarrow}{M}(M-3).
\end{equation}
For the  spectrum of the operator $i\hbar\frac{\partial}{\partial{t}}$ we have $E=\hbar\omega(2n+l+M)$.
 
 The eigensolutions  of the operator $\hat{P_0}(\rho,{\bf n})$
\begin{equation}
\label{1.87}
{\lbrace}H(\rho,{\bf n})+\frac{m{\omega}^2}{2}\left(\rho-i\frac{\hbar}{mc}\right)^2e^{-i\frac{\hbar}{mc}{\partial}_{\rho}}{\rbrace}{\xi}=\hbar\omega(2n+l+M){\xi}
\end{equation}
are $(\widetilde\rho=\frac{mc\rho}{\hbar}$, $M=3/2+\sqrt{1/4+(\frac{mc^2}{\hbar\omega})^2}$ $)$
\begin{equation}
\label{1.88}
{\xi}^{M}_{nlm}=c^{M}_{nl}(\frac{\hbar\omega}{mc^2})^{-i\widetilde\rho}\Gamma(M-1-i\widetilde\rho)i^{l+1}\frac{\Gamma[l+1+i\widetilde\rho)}{\widetilde\rho\Gamma(i\widetilde\rho)}{\cal P}^{M}_{n,l}({\widetilde\rho})Y_{lm}(\theta_1,\varphi_1),
\end{equation}
where the polynomials  ${\cal P}^{M}_{n,l}({\widetilde\rho})$ may be constructed with the help of the recursion relations 
\begin{equation}
\label{1.89}
b_n{\cal P}^{M}_{n+1,l}({\widetilde\rho}^2)+d_n{\cal P}^{M}_{n-1,l}({\widetilde\rho}^2)=[d_n+b_n-{\widetilde\rho}^2-l^2]{\cal P}^{M}_{n,l}({\widetilde\rho}^2),
\end{equation}
\begin{equation}
\label{1.90}
b_n=(n+1)(n+M-1/2),{\quad}d_n=(n+l+M-1)(n+l+1/2),
\end{equation}
\begin{equation}
\label{1.91}
{\cal P}^{M}_{-1,l}({\widetilde\rho}^2)=0,{\quad}{\cal P}^{M}_{0,l}({\widetilde\rho}^2)=1.
\end{equation}
The normalisation constants are 
\begin{equation}
\label{1.92}
c^{M}_{nl}=\frac{1}{\Gamma(M-1/2)}\left[{\frac{2n!\Gamma(n+M+1/2)}{\Gamma(n+l+3/2)\Gamma(n+M+l)}}\right]^{1/2}.
\end{equation}

For the function $\Phi(\rho,{\bf n};t,{\bf x})$, we have
\begin{equation}
\label{1.93}
{\Phi}^{M}=\sum_{n=0,l=0}^{\infty}{\sum_{m=-l}^{m=l}}{\psi}^{M}_{nlm}{\xi}^{M}_{nlm}.
\end{equation}
From  
\begin{equation}
\label{1.94}
C(\rho,{\bf n}){\Phi}^{M}=C(\tau,\sigma,\theta_2,\varphi_2){\Phi}^{M}
\end{equation}
 we obtain a  relation like (\ref{1.44}),
\begin{equation}
\label{1.95}
\omega=\frac{mc^2}{\hbar\sqrt{(M-1)(M-2))}}.
\end{equation}
The final expression for the energy spectrum  takes the form
\begin{equation}
\label{1.96}
E_{nlM}=\frac{mc^2}{\sqrt{(M-1)(M-2)}}(2n+l+M).
\end{equation} 
 For the energy levels $E_{00M}$ one gets $E_{00M\rightarrow\infty}\to{mc^2}$. There is no zero-point energy.
 
For the AdS radius $\kappa=c/{\omega}$, we have $\kappa=\sqrt{(M-1)(M-2)}$ $\hbar/mc$.

\section{Conclusion}
In this paper we have shown that  the generalized Schr\"odinger (GS) picture may be used to describe  the motion of a relativistic particle in an external field. For the harmonic oscillator potential we found that the spacetime independent operators in the equations of states induce in a natural way the operators which are related to Killing vectors of the AdS space. The problem of determining  the  Hamilton operator of the particle in the external fields based  on  choosing  the coordinate system in this space. We found that the oscillator frequency and the AdS radius are discrete. We have shown that in the relativistic region there is no zero-point energy. We  constructed propagators which  describe  extended  relativistic and nonrelativistic harmonic oscillators in the GS picture.
\section{Appendix A}
In the momentum representation 
\begin{equation}
\label{1.97}
p_0={mc^2}{\cosh}{\chi},{\quad}p=mc\sinh\chi,\quad{\chi}=-\ln[(p_0-cp)/mc^2],
\end{equation}
the  operators $\hat{P_0}$ and $\hat{P}_1$ are 
\begin{equation}
\label{1.98}
\hat{P_0}(\chi)=mc^2\left[\cosh\chi-\frac{1}{2}\left(\frac{\hbar\omega}{mc^2}\right)^2{e}^{\chi}\left\lbrace\frac{d^2}{d{\chi}^2}+\frac{d}{d\chi}\right\rbrace\right],
\end{equation}
\begin{equation}
\label{1.99}
\hat{P}_1(\chi)=mc\left[\sinh\chi-\frac{1}{2}\left(\frac{\hbar\omega}{mc^2}\right)^2{e}^{\chi}\left\lbrace\frac{d^2}{d{\chi}^2}+\frac{d}{d\chi}\right\rbrace\right].
\end{equation}
These operators  satisfy the commutation relations
\begin{equation}
\label{1.100}
[{N(\chi)},\hat{P}_1(\chi)]=i\frac{\hbar}{mc^2}\hat{P_0}(\chi),\quad[\hat{P}_1(\chi),\hat{P_0}(\chi)]=-i\hbar{m}{\omega}^2{N(\chi)},
\end{equation}
\begin{equation}
\label{1.101}
[\hat{P_0}(\chi),{N(\chi)}]=-i\frac{\hbar}{m}\hat{P}_1(\chi),
\end{equation}
where
\begin{equation}
\label{1.102}
N(\chi)=i\frac{\hbar}{mc}\frac{\partial}{\partial{\chi}}.
\end{equation}

\section{Appendix B}
The functions  ${v}_n(\rho)$ in (\ref{1.37}) are $(\nu=\mu+1/2)$
\begin{equation}
\label{1.103}
{v}_n(\rho)=c_n\left(\frac{\hbar\omega}{mc^2}\right)^{-i\frac{mc\rho}{\hbar}}\Gamma\left(\nu-i\frac{mc\rho}{\hbar}\right)_2F_1\left(-n,\nu-i\frac{mc}{\hbar}\rho;2\nu;2\right)
\end{equation}
where 
\begin{equation}
\label{1.104}
c_n=(-i)^{n}2^{\nu}\sqrt{\frac{\Gamma(2\nu+n)}{2\pi{n!}{[\Gamma}(2\nu)]^{2}}}.
\end{equation}
The functions $f_n(t,x)$ in (\ref{1.37}) may be written in the form  (${\lambda}_n=\mu +1/2+n$)
\begin{equation}
\label{1.105}
f_n(t,x)=e^{-i{E_n}t/\hbar}\left(1+\frac{{\omega}^2{x}^2}{c^2}\right)^{-\frac{{\lambda}_n}{2}}\sqrt{\frac{\Gamma(2\mu+1)\Gamma(\mu+\frac{1}{2})}{n!\Gamma(2\mu+1+n)\sqrt\pi\Gamma(\mu+1)}}{\varrho}_n(x).
\end{equation}
Here, for the polynomials  ${\varrho}_n$ we have the relations $(l=c/\omega$,  $y=x/l$, ${\varrho}_0(y)=1$,  $n=1,2,3...$)
\begin{equation}
\label{1.106}
{\varrho}_{n}(y)=[2{\lambda}_{n-1}{y}-(1+y^2){\partial}_{y}]{\varrho}_{n-1}(y).
\end{equation}
For $n=1,2,3$ we have
\begin{equation}
\label{1.107}
{\varrho}_1(y)=2{\lambda}_0{y},{\quad}{\varrho}_2(y)=2{\lambda}_0[(2\lambda_0+1)y^2-1],
\end{equation}
\begin{equation}
\label{1.108}
{\varrho}_3(y)=4{\lambda}_0({\lambda}_0+1)y[(2\lambda_0+1)y^2-3].
\end{equation}

\end{document}